%% file: Implicit_Sensor-based_Authentication_of_Smartphone_Users_with_Smartwatch.tex
\documentclass{sig-alternate-05-2015}

\usepackage{graphicx}
\usepackage{epstopdf}
\usepackage{subfigure}
\usepackage{bm}
\usepackage{makecell}
\usepackage{enumerate}
\usepackage{amsmath}
\newcommand{\tabincell}[2]{\begin{tabular}{@{}#1@{}}#2\end{tabular}}%

\begin{document}

\CopyrightYear{2016} 
\setcopyright{acmcopyright}
\conferenceinfo{HASP 2016,}{June 18 2016, , }
\isbn{978-1-4503-4769-3/16/06}\acmPrice{\$15.00}
\doi{http://dx.doi.org/10.1145/2948618.2948627}

\title{Implicit Sensor-based Authentication of Smartphone Users with Smartwatch}

%
%
%
%
%

\numberofauthors{2} 
%
\author{
%
%
\alignauthor
Wei-Han Lee\\
       \affaddr{Princeton University}\\
       \email{weihanl@princeton.edu}
\alignauthor
Ruby Lee\\
       \affaddr{Princeton University}\\
       \email{rblee@princeton.edu}
}



\maketitle
\input{abstract}

%
%

%
%

%
%
\printccsdesc



\input{introduction}
\input{related_work}
\input{system}
\input{algorithm}
\input{evaluation}

\input{conclusion}
\vspace{-1em}
\section{Acknowledgment}
This work was supported in part by NSF CNS 1218817.
\vspace{-1em}
\bibliographystyle{abbrv}
\bibliography{v3}  

\end{document}

%% file: abstract.tex
\begin{abstract}
Smartphones are now frequently used by end-users as the portals to cloud-based services, and smartphones are easily stolen or co-opted by an attacker. Beyond the initial log-in mechanism, it is highly desirable to re-authenticate end-users who are continuing to access security-critical services and data, whether in the cloud or in the smartphone. But attackers who have gained access to a logged-in smartphone have no incentive to re-authenticate, so this must be done in an automatic, non-bypassable way. 
Hence, this paper proposes a novel authentication system, iAuth, for implicit, continuous authentication of the end-user based on his or her behavioral characteristics, by leveraging the sensors already ubiquitously built into smartphones. We design a system that gives accurate authentication using machine learning and sensor data from multiple mobile devices. Our system can achieve $92.1\%$ authentication accuracy with negligible system overhead and less than $2\%$ battery consumption. 
\end{abstract}

%% file: introduction.tex
\section{Introduction}
We consider two usage scenarios in this paper: attackers accessing sensitive cloud-based services and data through a smartphone, and attackers accessing sensitive data stored in the smartphone itself.

Public clouds offer elastic and inexpensive computing and storage resources to both companies and individuals. Cloud customers can lease computing resources, like Virtual Machines, from cloud providers to provide web-based services to their own customers - who are referred to as the end-users. 

Past work on protecting a cloud customers' Virtual Machines tended to focus on attacks within the cloud from malicious Virtual Machines that are co-tenants on the same server, or from compromised Virtual Machine Monitors, or from network adversaries \cite{szefer2011eliminating,liu2010new}. However, end-users can also pose serious security threats.

Consider the increasingly common situation of accessing cloud-based services and data through a smartphone. Users register accounts for these services. Then they login to their accounts from their smartphones and use these cloud services. However, after log-in, the user may leave her smartphone unattended or it may be co-opted by an attacker, and now the attacker has legitimate access to the cloud-based services and data, or the sensitive data stored in the smartphone itself. Ideally, smartphone users should re-autheticate themselves, but this is inconvenient for legitimate users and attackers have no incentive to "re-authenticate". This paper addresses how re-authentication can be done conveniently, without explicit user participation, for smartphone users.

In the second scenario, smartphones themselves store private, sensitive and secret information related to our daily lives. We do not want these accessible to an attacker who has stolen the device, or has temporary access to it.

To protect cloud-based services and data from adversaries who masquerade as legitimate end-users, we propose a secure and usable re-authentication system, which is both \textit{implicit} and \textit{continuous}. An implicit authentication method does not rely on the direct involvement of the user, but is closely related to her behavior or living environment. This is more convenient than having to re-enter passwords. A continuous re-authentication method should keep authenticating the user, in addition to the initial login authentication. This can detect an adversary once he gets control of the smartphone and can prevent him from accessing sensitive data or services via smartphones, or inside smartphones.
Our system, called iAuth, can protect cloud-based services and data from attackers who masquerade as end-users, to enhance any security already provided by the cloud providers to cloud customers. iAuth can also help protect the critical information stored in the smartphone. The smartphone stores private and confidential information, which should not be accessible to an adversary who steals or somehow gets temporary access to the smartphone. iAuth is able to identify the adversary and restrict the adversary's access to sensitive information, even when the smartphone has no network services.

iAuth exploits one of the most important differences between personal computers and smartphones: a variety of sensors built into the smartphone, such as accelerometer, gyroscope, magnetometer and ambient light, etc. iAuth also exploits the increasing number of wearable devices with Bluetooth connectivity and multiple sensors, e.g., smartwatches. It is designed based on the fact that sensor measurements within the smartphones and wearable devices can reflect users' behavioral patterns, thus achieving highly accurate user authentication.

We propose some new techniques in iAuth to overcome the limitations posed by past smartphone authentication methods. 
(1) Some past work had high authentication errors~\cite{cc2, mantyjarvi2005icassp}. We combined a smartwatch with a smartphone to improve the authentication accuracy. However, it is challenging to combine multiple devices since they usually contain a large amount of noise that may influence the authentication accuracy if not handled properly. We successfully address this problem by utilizing both time and frequency information of the sensors' data from multiple devices. 
(2) Past approaches require a long time to learn a user's behavior or detect attacks~\cite{cc3,cc4}. We use sophisticated machine learning algorithms in iAuth, taking only $13$ milliseconds to identify any unauthorized accesses to the devices. This can block the adversaries before they steal any useful information. 
(3) Some past work only do one-time authentication~\cite{de2012touch}, while iAuth enables continuous authentication as a background service, when the user is using a smartphone. 
(4) Our system incurs rather low CPU and memory overhead, and only costs $2\%$ additional battery power, on modern smartphones. We believe such lightweight properties would make iAuth an attractive system for continuous authentication in real world applications.
Our key contributions are:

\begin{enumerate}[$\bullet$]
\item Design of an implicit authentication system, iAuth, by combining a user's sensor information recorded in the smartphone and wearable devices. Our system continuously monitors the user's behavior and authenticates the user in an accurate, efficient, and stealthy manner.
\item An efficient and low-overhead use of sensor measurements as behavioral patterns in both time and frequency domains, and an efficient machine learning classifier, for low overhead authentication.  
\item Experimental results to show that our approach can achieve high authentication accuracy up to $92.1\%$.
\end{enumerate}

%% file: related_work.tex
\section{Related Work}\label{sec:related}
Traditional authentication approaches are based on possession of secret information, such as passwords. Also, physiological biometrics based approaches make use of distinct personal features, such as fingerprints or iris patterns. Recently, behavior-based authentication utilize the distinct behavior of users, e.g., gaits and gestures.

Currently, there are many different physiological biometrics for authentication, such as face patterns, fingerprints \cite{hong1998pami}, and iris patterns \cite{qi2008iris}. However, physiology-based authentication requires user participation in the authentication. For example, fingerprint authentication needs the user to put his finger on the fingerprint scanner. Hence, these physiology-based approaches requiring user compliance can not achieve continuous and implicit authentication.

Behavior-based authentication assumes that people have distinct stable patterns for a certain behavior, such as hand-writing pattern \cite{cc5,shahzad2013secure}, gait \cite{cc7} and GPS patterns \cite{cc4}. Behavior-based authentication exploits users' behavioral patterns to authenticate a user's identity. Below we review past work in this area that specifically use sensors built into smartphones.

{\bf Smartphone Authentication with Sensors.}
Kayacik et al. \cite{cc3} proposed a lightweight, and temporally \& spatially aware user behavioral model for user authentication based on both hard and soft sensors. They showed that the attacker can be detected in $717$ seconds. However, they did not quantitatively show their authentication performance.
SenSec \cite{cc2} constantly collects data from the accelerometer, gyroscope and magnetometer, to construct gesture models while the user is using the device.
GPS sensors are used in \cite{cc4} to demonstrate that the system could detect abnormal activities (e.g., a phone being stolen) by analyzing a user's location history. 
Shahzad et al. \cite{shahzad2013secure} and Trojahn et al. \cite{cc5} developed a mixture of a keystroke-based and a handwriting-based method to realize authentication through the screen sensor.  
Li et al.\cite{cc6} exploited five basic movements (sliding up, down, right, left and tapping) and their related combinations as the user's behavioral pattern features, to perform authentication on smartphone.
Nickel et al. \cite{cc7} used accelerometer-based behavior recognition to authenticate a smartphone user through the $k$-NN algorithm. 
Lee et al. \cite{lee2015icissp,lee2015implicit} showed that using more sensors can improve authentication performance. They monitored users' living patterns and utilized SVM as a classifier for user authentication. 
Our iAuth system has better authentication accuracy (around $92\%$) with lower complexity than previous methods. 

Riva et al. \cite{riva2012progressive} built a prototype to use face recognition, proximity, phone placement, and voice recognition to progressively authenticate a user. However, their objective is to decide when to authenticate the user and is thus orthogonal to our setting. Furthermore, their scheme requires access to sensors that need users' permissions, limiting their applications for implicit authentication.

{\bf Authentication with Wearable Devices.}
Recently, wearable devices have emerged in our daily lives. However, limited research has been done on authenticating users by these wearable devices. In \cite{mare2014zebra}, Mare et al. proposed ZEBRA which is a bilateral recurring authentication method. The signals sent from a bracelet worn on the user's wrist are correlated with the terminal's operations to confirm the continued presence of the user if the two movements correlate according to a few coarse-grained actions. To the best of our knowledge, there is no smartphone authentication research proposed in the literature that combines a wearable smartwatch with a smartphone to authenticate a user, as we do.

%% file: system.tex
\section{Threat model and Assumptions}
We consider an attacker who has physical access to the smartphone and aims to steal sensitive information or cloud services accessed via the smartphone. We focus on this type of attacks based on two observations. First, compared to traditional computing devices (e.g., PCs), smartphones are small and easily lost or stolen. Besides, users sometimes leave their smartphones for a while, which give attackers opportunity to access the critical information. Second, current authentication methods (e.g., passwords) can be misused by users. For instance, a lot of users do not set the password since it is inconvenient to input it every time, and other users choose weak passwords. Our threat model assumes passwords are vulnerable.

We assume the smartphone functions correctly: the sensors in the smartphone are trusted to provide accurate data. Some secure part of the system software is able to lock the smartphone, or deny access to security-critical resources, once the user does not pass the authentication test. The integrity of the iAuth app in the smartphone is verified and secure, so the attacker cannot bypass the authentication by compromising the iAuth software.

The smartwatches have built-in sensors, e.g., accelerometer and gyroscope. They are also equipped with a wireless radio (e.g., Bluetooth) to communicate with smartphones. We assume the communication between the smartwatch and smartphone is secure. We assume each smartwatch (and smartphone) is associated with one owner and users do not share their smartwatches (and smartphones). This association can be implemented using a variety of approaches. For instance, a user may be required to enter a validation code when she puts the smartwatch on to activate it, and the smartwatch would deactivate when it is removed from the wrist, which can be detected through the built-in sensors such as accelerometer, or after a certain period of time. Our system works if only the smartphone is present, but we will show that it works better if the smartwatch is also present.

iAuth uses backend services for computing and training authentication models. These services are located in remote cloud servers (Authentication Servers), which are assumed to be trusted. The confidentiality and integrity of authentication servers are protected so the attackers cannot steal or modify the users' sensor data stored in the server's database. The threat model also assumes that attackers do not succeed in denial of service attacks on the authentication services. We also assume the communication channels between the smartphones and authentication servers are secure, e.g., SSL is used for communication between the smartphone and the authentication server in the cloud. Actual authentication, after training, is done in the smartphone itself, without the need to access the cloud or networks.
 
\section{System Design} \label{sec:design}

\subsection{Architecture Overview}\label{archi}
\begin{figure}[t]
\centerline{\mbox{\includegraphics[width=1.1\linewidth]{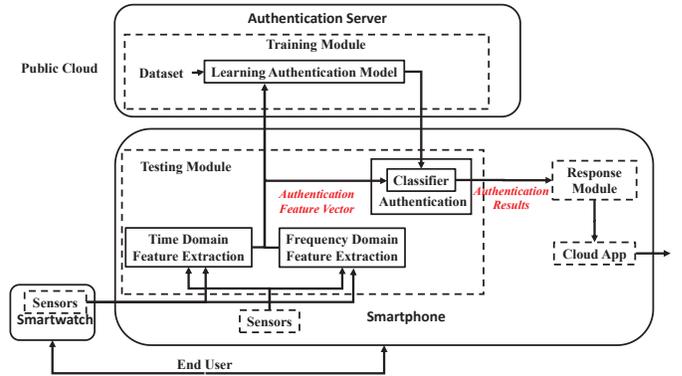}}}
\caption{{\footnotesize iAuth architecture including the cloud-based training module and smartphone-based testing module}}
\label{fig:shadow}
\end{figure}

Figure~\ref{fig:shadow} shows the iAuth architecture. It includes three hardware devices: the wearable device (e.g., smartwatch), the smartphone, and the authentication server in the cloud.

\subsubsection{Wearable IoT device.} 
In iAuth, we consider a two-device authentication configuration, which includes a mobile smartphone and a user-owned wearable device. We use a smartwatch as an example, but other types of wearable devices, e.g., body sensors, can also be applied to iAuth. iAuth authentication is designed for implicit authentication on the smartphone, where the smartwatch serves as important auxiliary information for improving authentication accuracy. The smartwatch keeps monitoring a user's raw sensors' data and sends the information to the smartphone via Bluetooth. 

\subsubsection{Smartphone.}
The smartphone also monitors the user's sensor data. It runs the authentication testing module as a background service. In the testing module, sensor data from the smartphone and smartwatch are sent to the feature extraction components, in both the time domain and the frequency domain, where fine-grained time-frequency features are extracted to form the authentication feature vector. This is fed into the authentication component.

The classification algorithm used in our work is the kernel ridge regression (KRR) algorithm \cite{suykens2002krr}, but other machine learning algorithms can also be used. The authentication component consists of a classifier, which is used to authenticate the user based on the authentication feature vector.

When the classifier generates the authentication results, it sends these results to the Response Module. If the authentication results indicate the user is legitimate, then the Response Module will allow the user to use the cloud apps to access the critical data or cloud services in the app server. Otherwise, the Response Module can either lock the smartphone or refuse accesses to the security-critical data, or perform further checking. If the legitimate user is misclassified, in order to unlock the smartphone, several possible responses can be implemented, depending on the situation and security requirements. For example, the legitimate user must explicitly re-authenticate by using a biometric that may have been required for initial log-in, e.g., a fingerprint. The legitimate user is motivated to unlock his device, whereas the attacker does not want to use his fingerprint because it will leave a trace to his identity. iAuth architecture allows such explicit unlocking mechanisms, but is not restricted to one such mechanism.

\subsubsection{Authentication Server.} 
iAuth includes a training module deployed in the Authentication Server in the cloud. This provides efficient computation and enables the training data set to use sensor feature vectors of other enrolled smartphone users. When a legitimate user first enrolls in the system, the system keeps collecting the legitimate user's authentication feature vectors for training the authentication model. Our system deploys a trusted Authentication cloud server to collect sensors' data from all the participating legitimate users. To protect a legitimate user's privacy, the entire users' data are anonymized \cite{liu2016linkmirage, liu2016ddp}. In this way, a user's training module can use other users' sensor data but has no way to know the other users' identities. The training module uses the legitimate user's authentication feature vectors and other people's authentication feature vectors in the training algorithm to obtain the authentication model. After training, the authentication model is downloaded to the smartphone. The training module does not participate in the authentication testing process and is only needed for retraining when the device recognizes a user's behavioral drift, which is done online and automatically. Therefore, our system does not require continuous communication between the smartphone and the Authentication Server.

\subsection{System Operation}
iAuth stems from the observation that behavioral patterns are different from person to person, when using smartphones and smartwatches.

There are two phases for learning and classifying the user's behavioral pattern: \textit{enrollment phase} and \textit{continuous authentication phase}. iAuth learns a profile of the legitimate user in the enrollment phase and then authenticates the user in the continuous authentication phase.

\noindent{\bf{Enrollment Phase}}: Initially, the system must be trained in an enrollment phase. When the users want to use the apps in the smartphone to access sensitive data or cloud services, the system starts to monitor the sensors and extract particular features from the sensors' data. This process continues and the data should be stored in a protected buffer in the smartphone until the distribution of the collected features converges to an equilibrium, which means the size of data can provide enough information to build a user's profile. This is about $800$ measurements for our method, as shown in Section~\ref{datasize}. At this time, one can assume that 1) the user got used to her device and her device-specific `sensor-behavior' no longer changes, and 2) the system has observed sufficient information to have a stable estimate of the true underlying behavioral pattern of that user. The system can now train the authentication classifier and switch to the continuous authentication phase.

\noindent{\bf{Continuous Authentication Phase}}: Once the authentication classifier is trained and sent to the smartphone, the smartphone can start the authentication phase. This is done only in the smartphone, so network availability is not required. Based on the sensor data, the authentication classifier decides whether these sensors' data are coming from the legitimate user.

\noindent{\bf{Post-Authentication}}: If the authentication feature vector is authenticated as coming from the legitimate user, this testing passes and the user can keep accessing the sensitive data in the smartphone or in the cloud via the smartphone. When an attacker tries to access a smartphone of a legitimate user, the system automatically de-authenticates him. Once iAuth decides that the smartphone is now being used by someone other than the legitimate user, the system can perform defensive responses as described earlier. Similarly, if the legitimate user is misclassified, several mechanisms for re-instating her are possible, such as two-channel or multi-factor authentication, or requiring an explicit login again, possibly with a biometric, to unlock the system.

\subsection{Security Protections}
In this subsection, we describe the security protections needed for the iAuth system.

\noindent{\bf{Protecting data in transit}}. Sensitive data are transmitted between smartwatches, smartphones and cloud servers. Secure communications protocols are exploited to provide confidentiality and integrity protection against network adversaries. For instance, an initialization key is exchanged when the smartwatch is paired with the smartphone using Bluetooth. New keys derived from this key can also be used to encrypt and hash the raw data transmitting between smartwatch and smartphone via Bluetooth. The communication channels between smartphones and cloud servers are protected by SSL/TLS protocols~\cite{dierks2008transport}. These network transmissions between a smartphone and the cloud are minimized, since iAuth saves the latest $n$ sensor measurements in a trusted buffer (e.g., using ARM Trustzone), and sends these in a batch to the cloud only on initial training. $n$ is the number of data samples needed (see Sections~\ref{datasize})
 
\noindent{\bf{Protecting data at rest (i.e., in storage)}}. When the data are stored in the smartphones, or cloud servers, cryptographic encryption and hashing operations are used to prevent the attackers from stealing or modifying data. 

\noindent{\bf{Protecting data and code at runtime.}}
The smartphone and Authentication Server must also provide a secure environment for running iAuth authentication testing and training code, and using sensitive sensor measurements collected from different users and devices. Since most smartphones use ARM processors, smartphones can exploit the ARM TrustZone feature to place the authentication Testing Module in the Secure World and isolate it from other apps in the Normal World. 
The wrap-around buffer for collecting the latest sensor measurements, discussed above, can also be in Trustzone's secure storage in the Secure World.
Since cloud servers tend to use Intel processors, they will soon have access to Intel Software Guard eXtensions (SGX) \cite{hoekstra2013using,anati2013innovative,mckeen2013innovative}. Hence, the trusted Authentication Server can set up secure enclaves for the training and retraining modules for iAuth, and for securely accessing and using sensitive behavioral measurements from many smartphone users.
Alternatively, some method of securely protecting trusted application code, even from potentially compromised Operating Systems, is needed. For example, Bastion secure trusted environments can be provided \cite{champagne2010scalable} for protection equivalent to SGX secure enclaves.

%% file: algorithm.tex
\section{Authentication Algorithms}
\subsection{Sensor Selection}
We select the following two sensors: accelerometer and gyroscope~\cite{google}, in smartphones because: (1) They are ubiquitously built into current smartphones. (2) These two sensors also represent different information about the user's behavior. The accelerometer records the motion patterns of a user such as how she walks \cite{cc7}. The gyroscope records fine-grained motions of a user such as how she holds a smartphone \cite{cc8}. (3) These sensors do not need the user's permissions, making them useful for continuous background monitoring in implicit authentication. (4) The sensor data itself does not contain information usually considered privacy sensitive, like GPS locations, screen data and voice. 

We use a wearable device like a smartwatch to provide further information on user behavior to enhance user authentication accuracy. We use the same sensors in the smartwatches for the same reasons discussed above. We will show that even if the same type of sensors are used on the smartphone and the smartwatch, but on different parts of the body, they record different aspects of a user's behavior.

Although the proximity of the smartwatch to the smartphone can be used as a simple second-factor authentication signal, this may be less secure if the attacker gains access to the victim's smartphone while the victim, wearing his smartwatch, is still within Bluetooth connectivity range. If the attacker gets  access to both the smartphone and the smartwatch, a simple proximity signal of the smartwatch will not help de-authenticate the attacker - smartwatch sensors would work better. Hence, we propose using sensors on the smartwatch (or other wearable) for improving smartphone user authentication.

\subsection{Authentication}\label{sec:authentication}

\subsubsection{Time Domain and Frequency Domain.}
We segment the signals of the sensors' data into a series of time windows. In each window, we extract features from the time domain and the frequency domain of the sensors' data collected during this time from the accelerometer and gyroscope.

We first compute the magnitude of each sensor data. For an accelerometer data sample $(x,y,z)$ at time $t$, the magnitude is $m=\sqrt{x^2+y^2+z^2}$.
We denote the magnitude signal of sensor $i$ in the $k$-th window as $S_i(k)$.

 In the time domain, we extract the \emph{mean}, \emph{variance}, \emph{max} and \emph{min} at each time window. Thus, the features for sensor $i$ in the $k$-th window can be represented as 
\begin{equation} 
 SP_{i}^t(k)=[mean(S_{i}(k)),~var(S_{i}(k)),~max(S_{i}(k)),~min(S_{i}(k))]
\end{equation} 
 We obtain the authentication feature vector in the time domain in the $k$-th window as 
\begin{equation}
Auth^{t}(k)=[SP^t(k),SW^t(k)]
\end{equation}
 where 
\begin{equation}
\begin{aligned}
SP^{t}(k) &= [SP_{acc}^{t}(k),SP_{gyro}^{t}(k)]\\
SW^{t}(k) &=[SW_{acc}^{t}(k),SW_{gyro}^{t}(k)]
\end{aligned}
\end{equation} 

We also implement the Discrete \emph{Fourier} transform (DFT) \cite{boashash2003time} to obtain the frequency domain information. In the frequency domain, we extract three features: (1) The \emph{amplitude of the first highest peak}, which represents the energy of the entire sensors' information within the window, (2) the \emph{frequency of the second highest peak}, which represents the main walk frequency, and (3) the \emph{amplitude of the second highest peak}, which corresponds to the energy of the sensors' information under this dominant periodicity. 
The feature vector of window $k$ in the frequency domain for sensor $i$, can be represented as 
\begin{equation}
SP_{i}^{f}(k)=[energy(S_{i}(k)),freq(S_{i}(k)),energy_{fre}(S_{i}(k))]
\end{equation}
We construct the authentication feature vector in the frequency domain as 
\begin{equation}
Auth^{f}(k)=[SP^f(k),SW^f(k)]
\end{equation}
 where 
\begin{equation}
\begin{aligned}
 SP^{f}(k) &=[SP_{acc}^{f}(k),SP_{gyro}^{f}(k)] \\ 
 SW^{f}(k) &=[SW_{acc}^{f}(k),SW_{gyro}^{f}(k)]
\end{aligned}
\end{equation} 

 Combining authentication features from both time and frequency domains, we have the whole authentication feature vector as 
 \begin{equation}
 Auth(k)=[Auth^t(k),Auth^{f}(k)]
 \end{equation} 
After we obtain the time-frequency feature vectors, we utilize a light-weight machine learning approach, kernel ridge regression (KRR) \cite{suykens2002krr}, to train the authentication models in the cloud for user authentication. Our experimental results in Section~\ref{sec:experiments} show the improved accuracy of including frequency domain features.

\begin{figure*}
\centering
\subfigure[Accelerometer $x$]{
\label{fig:acc1_context1}
\epsfig{file=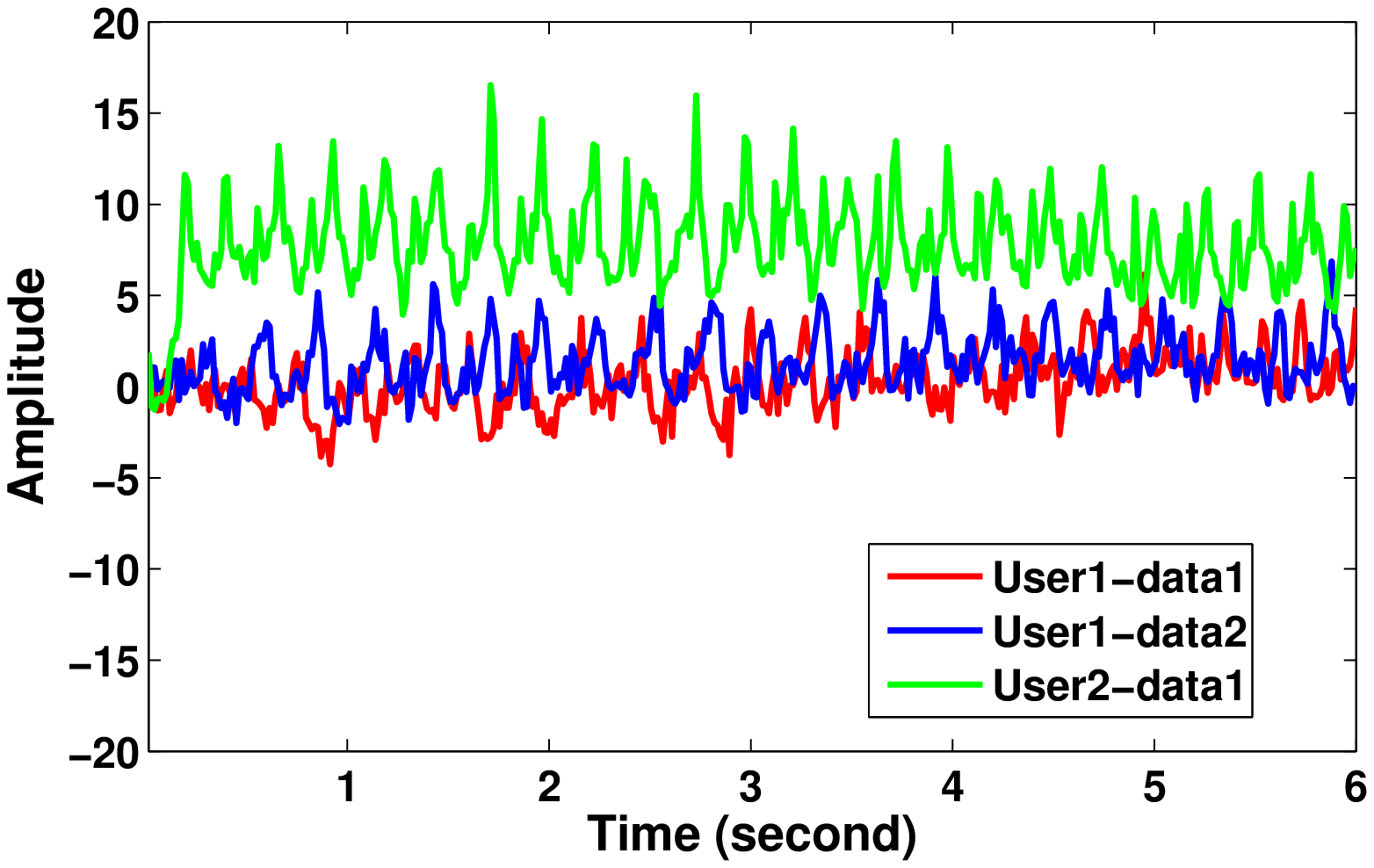, height=1.2 in, width=2.1 in}}
\subfigure[Accelerometer $y$]{
\label{fig:acc2_context1}
\epsfig{file=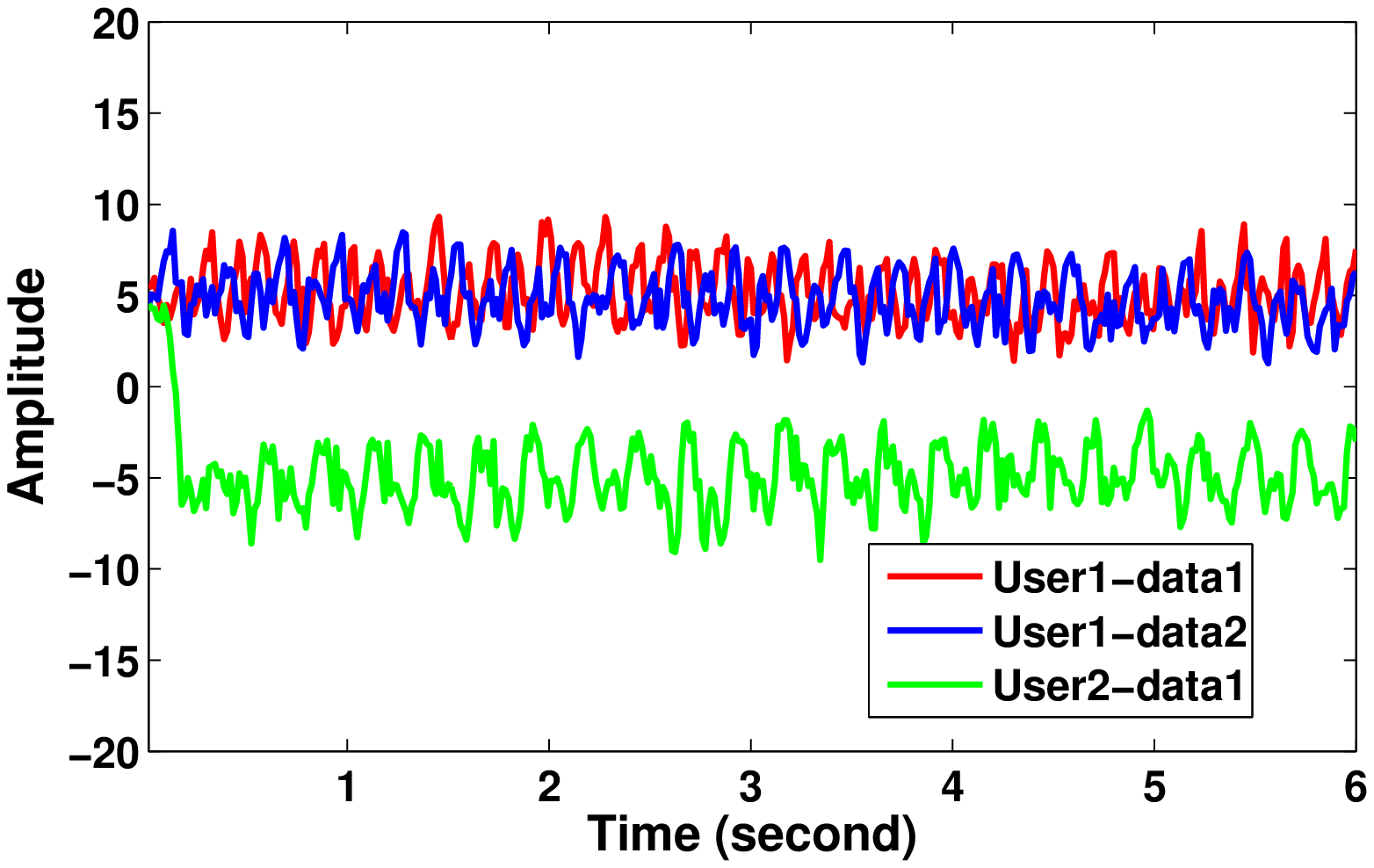, height=1.2 in, width=2.1 in}}
\subfigure[Accelerometer $z$]{
\label{fig:acc3_context1}
\epsfig{file=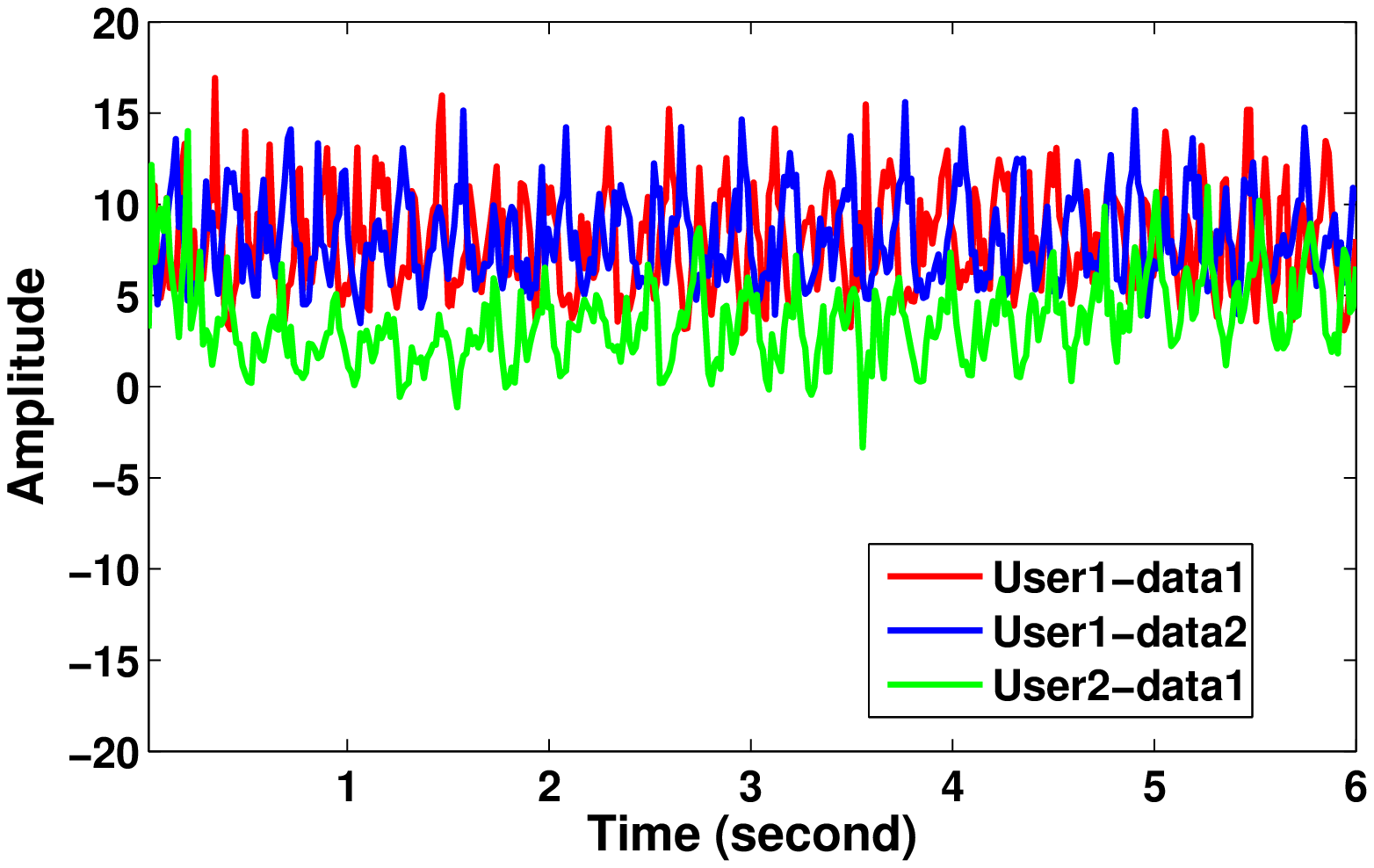, height=1.2 in, width=2.1 in}}
\subfigure[Gyroscope $x$]{
\label{fig:gyro1_context1}
\epsfig{file=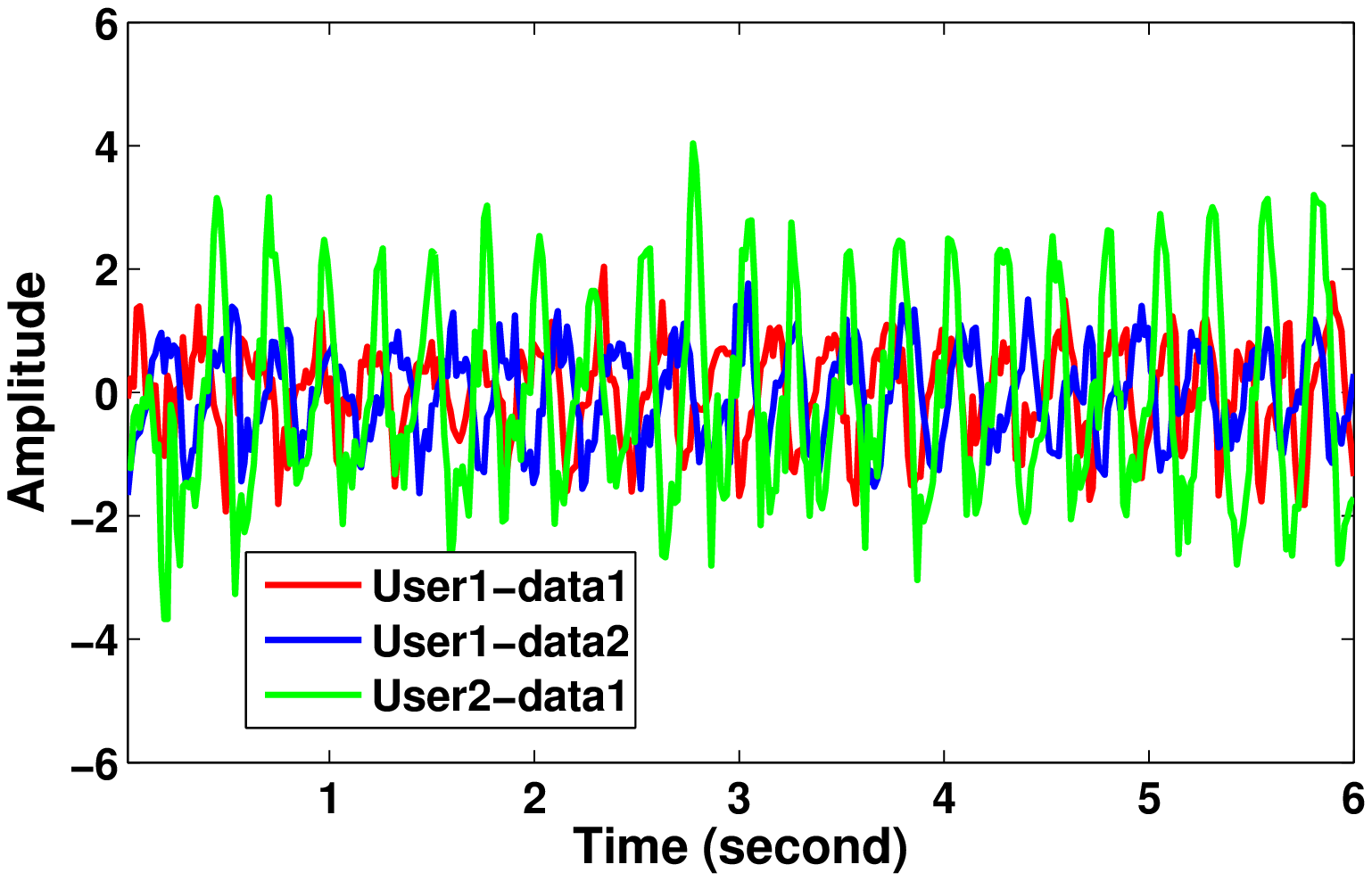, height=1.2 in, width=2.1 in}}
\subfigure[Gyroscope $y$]{
\label{fig:gyro2_context1}
\epsfig{file=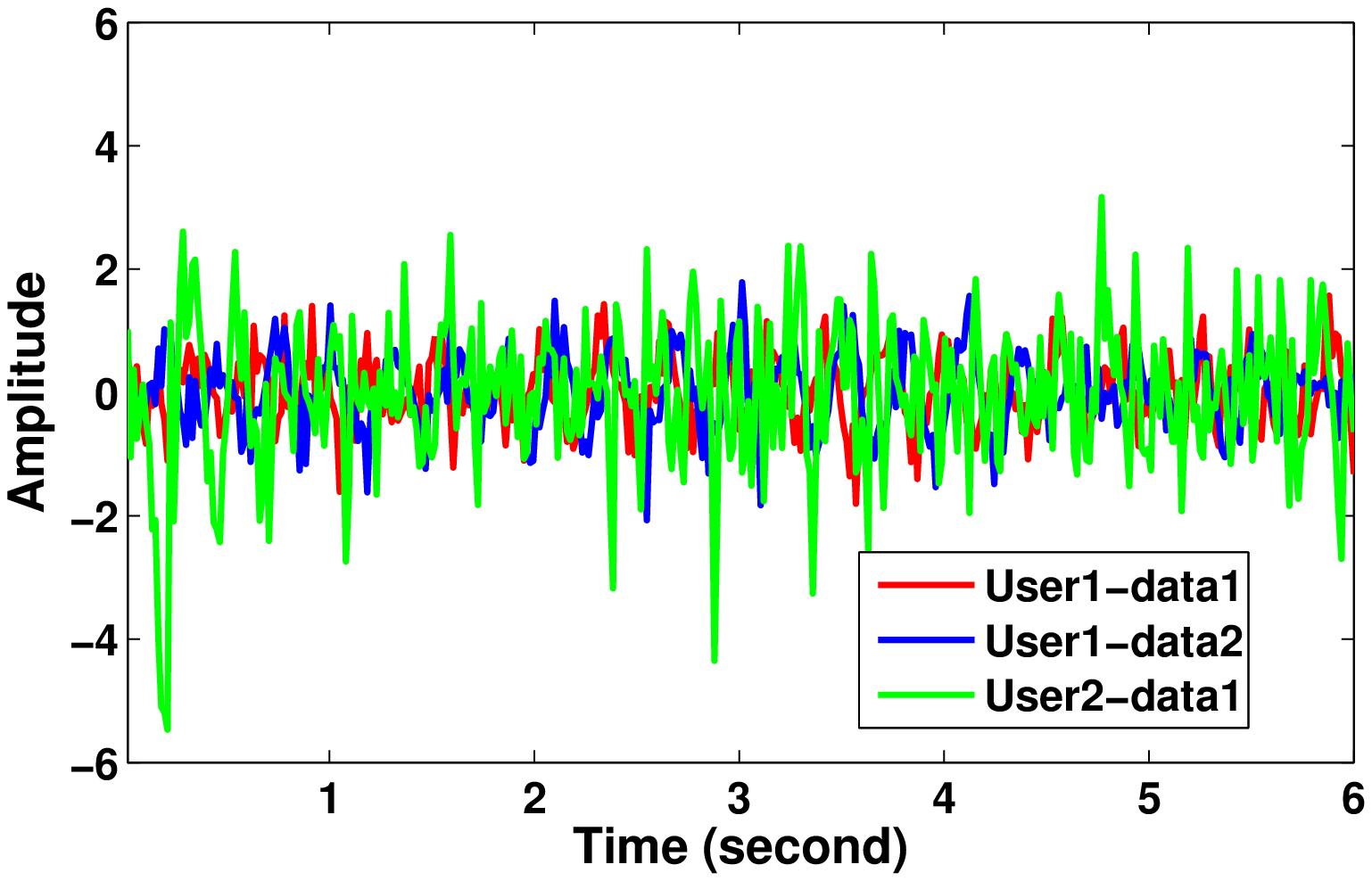, height=1.2 in, width=2.1 in}}
\subfigure[Gyroscope $z$]{
\label{fig:gyro3_context1}
\epsfig{file=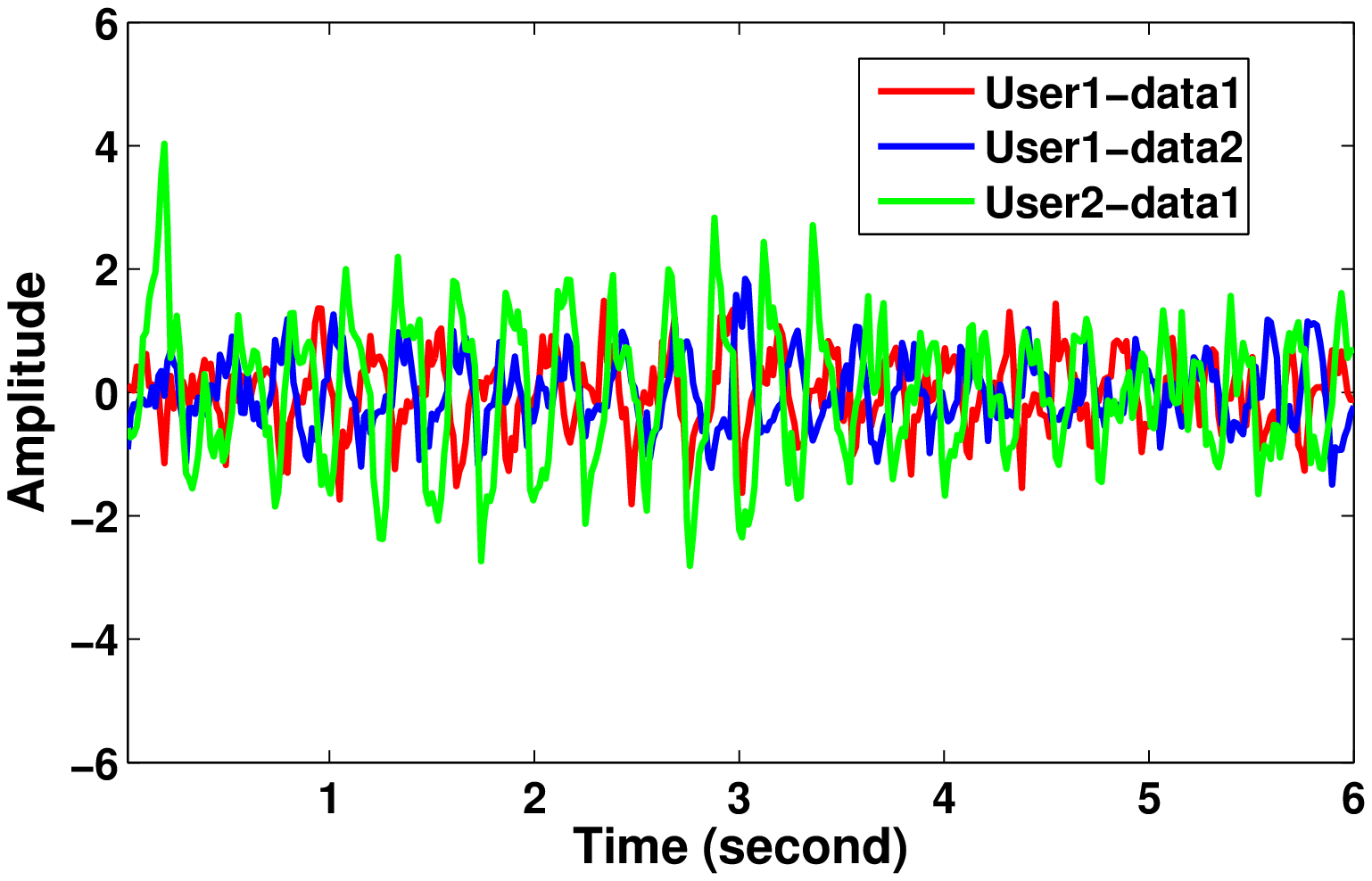, height=1.2 in, width=2.1 in}}
\caption{{\footnotesize The visualization of signals extracted from the accelerometer and gyroscope for three different axis. We randomly select two signals from the same user (red and blue (dark) lines) and a signal from another user (green (light) lines). We can see that the two sensor signals from the same user are more similar than that from different users.}}
\label{fig:visualization1}
\end{figure*}

\subsubsection{Kernel Ridge Regression (KRR)}\label{sec:krr}
Kernel ridge regressions (KRR) \cite{suykens2002krr} have been widely used for classification analysis. The advantage of KRR is that the computational complexity is much less than other machine learning methods, e.g., support vector machines (SVM) \cite{svm, lee2015icissp}. The goal of KRR is to learn a model that assigns the correct label to an unseen testing sample. This can be thought of as learning a function $f: X \rightarrow Y$ which maps each data $x$ to a label $y$. The optimal classifier can be obtained analytically according to
\begin{equation}\label{eq2}
\bm{w}^{*}=\mathrm{argmin}_{\bm{w}\in\mathbb{R}^d}\rho\|\bm{w}\|^2+\sum_{k=1}^N (\bm{w}^T {\bm{x}}_k -y_k)^2
\end{equation}
where $N$ is the data size and $\bm{x}^{M\times1}_k$ is the transpose of $Auth(k)$, and $M$ is the dimension of this authentication feature vector. Let $\bm{X}= [{\bm{x}}_1,{\bm{x}}_2, \cdots, {\bm{x}}_N]$ denote a $M\times N$ training data matrix. Let $\bm{y}= [{\bm{y}}_1,{\bm{y}}_2, \cdots, {\bm{y}}_N]$.
$\bm{\vec{\phi}(\bm{x_i})}$ denotes the kernel function, which maps the original data $\bm{x_i}$ into a higher-dimensional ($J$) space. In addition, we define $\bm{\Phi} = [\bm{\vec{\phi}(\bm{x_1})}\bm{\vec{\phi}(\bm{x_2})}\cdots \bm{\vec{\phi}(\bm{x_N})}] $ and $\bm{K}=\bm{\Phi}^T\bm{\Phi}$.
This objective function in Eq.~\ref{eq2} has an analytic optimal solution \cite{suykens2002krr} where
\begin{equation}\label{optimal}
\bm{w}^{*} =\bm{\Phi}[\bm{K}+\rho \bm{I_N}]^{-1} \bm{y}
\end{equation}
By utilizing certain matrix transformation properties, the computational complexity for computing the optimal $\bm{w}^{*}$ in Eq.~\ref{optimal} can be largely reduced from $O(N^{2.373})$ to $O(M^{2.373})$. Given that $N$ is abut $800$ samples and $M$ is $27$ for our $Auth(k)$ feature vector, this is a big reduction.

%% file: evaluation.tex
\section{Evaluation}\label{sec:experiments}
\subsection{Experiments setting} \label{sec:setting}
In order to evaluate the performance of our system, $20$ users are invited to take our smartphone and smartwatch for one to two weeks and use them in the same way that they use their personal smartphones and smartwatches in their daily lives. We collected sensor data from the accelerometer and gyroscope in a smartphone (Nexus 5) and a smartwatch (Moto 360) with a sampling rate of $50$ Hz. 

We perform two experiments. The first experiment is free-form usage of the smartphone and/or the smartwatch to determine the authentication parameter selection (Section \ref{sec:parameter}) and evaluate the authentication performance (Section \ref{sec:authenticationPerformance}). The second experiment tries to trick our system using masquerading attacks (Section \ref{sec:security}). Finally, we show the impact on the battery drainage (Section \ref{sec:overhead}). 

In our collected data, we used $10$-fold cross-validation to generate the training data and testing data sets for evaluating the authentication performance. To extensively investigate the performance of our system, we repeated such cross-validating mechanisms for $1000$ iterations and averaged the experimental results.

\begin{figure}[!t] \centering
\subfigure{
\includegraphics[width=2.1in,height=1.2in]{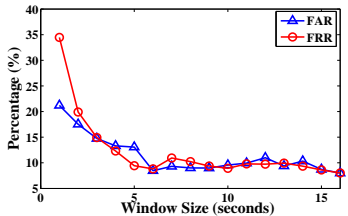}}
\caption{{\footnotesize FAR and FRR with different window sizes.}}
\label{fig:window_size} 
\end{figure}

\subsection{Confirming the intuition for iAuth}
While it may be hard to prove that sensor measurements can be used to differentiate users, we show some empirical results that confirm our intuition for our iAuth system by plotting the sensor data streams we collected from different users (as described in Section \ref{sec:setting}). Figures \ref{fig:visualization1} depicts the sensor streams corresponding to different sensor dimensions. 
For each sensor dimension, we randomly select two signal streams from the same user and one signal stream from another user for comparison.

In these figures, we observe that the sensor signals for the same user are more similar than those for different users, which lays the foundation for our authentication approach. Furthermore, by comparing the first row with the second row in Figure \ref{fig:visualization1}, we can see that the measurements from the accelerometer have more distinguishable characteristics than those of the gyroscope. For example, in Figure \ref{fig:acc1_context1}, the range of the acceleration of user 1 is completely separate from user 2, while the measurements of the gyroscope for the two users are overlapped to some extent as shown in Figure \ref{fig:gyro1_context1}.

\begin{figure}[!t] \centering
\subfigure{
\includegraphics[width=2.1 in,height=1.2in]{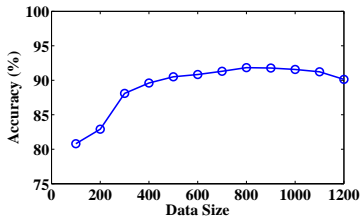}}
\caption{{\footnotesize Accuracy with different data sizes.}}
\label{fig:data_size} 
\end{figure}

\subsection{Authentication Parameter Selection}\label{sec:parameter}
In our first experiment, users were invited to take our smartphone and smartwatch for one to two weeks and use them under free-form real-use conditions. We let the participants use the devices in the same way that they use their personal devices in their daily lives.

There are two important parameters in the system, the window size and the size of the dataset. We first systematically investigate the performance of our approach under different values of these two parameters.

\subsubsection{Window Size}\label{windowsize}
The window size is an important system parameter, which determines the time that our system needs to perform an authentication, i.e., window size directly determines our system's authentication frequency.

We vary the window size from $1$ second to $16$ seconds. Given a window size, for each target user, we utilize $10$-fold cross-validation for training and testing. Here, we utilize the false rejection rate (FRR) and false acceptance rate (FAR) as metrics to evaluate the authentication accuracy of our system. FAR is the fraction of other users' data that are misclassified as the legitimate user's. FRR is the fraction of the legitimate user's data that are misclassified as other users' data. For security protection, a large FAR is more harmful than a large FRR. However, a large FRR would degrade the usage convenience. Therefore, we investigate the influence of the window size on FRR and FAR, in choosing a proper window size.

Figure~\ref{fig:window_size} shows that the FAR and FRR become stable when the window size is greater than $6$ seconds.

\subsubsection{Data Size}\label{datasize}
Another important system parameter is the size of the data set, which also affects the overall authentication accuracy because a larger training data set provides the system more information. According to our observations above, we set the window size to $6$ seconds. With the training set sizes ranging from $100$ to $1200$, we show the experimental results in Figure~\ref{fig:data_size}. We observe that the maximum accuracy happens when the data size is around $800$. The accuracy decreases after the training set size is larger than $800$ because a large training data set is likely to cause over-fitting in the machine learning algorithms so that the constructed training model would introduce more errors than expected.
\subsection{Authentication Performance}\label{sec:authenticationPerformance}
After setting up the system parameters of $6$ seconds window size and $800$ data size, we now show the overall authentication performance of our system in Table~\ref{table:krr}.

From Table~\ref{table:krr}, we have some interesting observations: (1) By using only the smartphone, our system can achieve $83.2\%$ authentication accuracy. (2) By combining the smartphone and smartwatch together, the authentication performance has a significant increase to $92.1\%$ authentication accuracy.

We measured the time for doing an authentication in our system to be less than $13$ milliseconds. Since the window size of our system is $6$ seconds, the time for doing an implicit authentication is roughly $6$ seconds. Therefore, our system can achieve good authentication performance within an acceptable time, making our system efficient and applicable in real world scenarios.

\begin{table}\small
\centering
\caption{{\footnotesize The FRR,FAR and accuracy of iAuth.}}
\begin{tabular}{|c|c|c|c|} \hline
 Device 	 & FRR  & FAR  & Accuracy\\ \hline
 Smartphone  &  $22.3\%$ & $13.4\%$	&  $83.2\%$ \\ \hline 
 Smartphone\& Smartwatch &  $8.3\%$  & $7.5 \%$	&  $92.1\%$ \\ \hline
\end{tabular}
\label{table:krr}
\end{table}

\subsection{Security Analysis}\label{sec:security}
We analyze our system's performance to defend against attacks, such as masquerading or mimicry attacks, in our next experiment. In this experiment, each subject was asked to mimic the victim user's behavior to the best of his ability as a malicious adversary. One user's behavior is recorded by a VCR and his/her model was built as the legitimate user. The other users were asked to watch the video and to try to mimic the legitimate user and pass the authentication testing. Such an attack is repeated $20$ times for each legitimate user and her corresponding `adversaries'.

Recall that the goal of iAuth is to prevent an attacker from getting access to the sensitive information stored in the cloud through the smartphone, or in the smartphone. iAuth achieves low FARs when attackers attempt to use the smartphone with their own behavioral patterns as shown in Figure \ref{fig:window_size} and Table \ref{table:krr}.

Now, we show that iAuth is even secure against the masquerading attacks where an adversary tries to mimic the user's behavior. Here, \textit{`secure'} means that the attacker cannot cheat the system via performing these spoofing attacks and the system should detect these attacks in a short time.
To evaluate this, we design a masquerading attack where the adversary not only knows the password but also observes and mimics the user's behavioral patterns. If the adversary succeeds in mimicking
the user's behavioral pattern, then iAuth will misidentify the adversary as the legitimate user and he can thus use the smartphone normally. 

In order to show the ability of iAuth in defending against the mimicry attacks, we counted the percentage of people (attackers) who were still using the smartphone without being de-authenticated by the system. Our experiments show that iAuth can detect $90\%$ of attackers in $18$ seconds on average. Also, iAuth identified all the adversaries within $24$ seconds. Therefore, iAuth performed well in recognizing the adversary who is launching the masquerading attack. 

Such experimental results also match with our analysis from a theoretical point of view. We assume the FAR at each time window is $p$, then the chance that the attacker can escape from detection in $n$ time windows is $p^n$. Based on our experimental results in Section \ref{sec:experiments}, our system can achieve $7.5\%$ FAR. Thus, within only three windows, the probability for the attacker escaping detection is $(7.5\%)^3 = 0.04\%$. Therefore, our iAuth shows good performance in defending against masquerading attacks.

\subsection{Smartphone Performance Overhead}\label{sec:overhead}
To demonstrate the applicability of our system in real world scenarios, we now evaluate the system overhead of iAuth on personal smartphones. Specifically, we analyze the CPU and memory overhead, and the battery consumption on the smartphone.

\subsubsection{CPU and Memory Overhead}
The testing module of iAuth in a smartphone runs as threads inside the smartphone system process. We develop an application to monitor the average CPU and memory utilization of the phone and watch while running the iAuth app which continuously requests sensor data at a rate of $50$ Hz on a Nexus 5 smartphone and a Moto 360 smartwatch. The CPU utilization is $4\%$ on average and never exceeds $6\%$. The CPU utilization (and hence energy consumption) will scale with the sampling rate. The memory utilization is $3$ MB on average. Thus, we believe that the overhead of iAuth is small enough to have negligible effect on overall smartphone performance.

\begin{table}[!t]\small
\centering
\caption{{\footnotesize Power consumption under four different scenarios.}}
\begin{tabular}{|l|c|} \hline
Scenario & \tabincell{c}{Power Consumption} \\ \hline
1) Phone locked, iAuth off & $2.8\%$ \\ \hline
2) Phone locked, iAuth on & $4.6\%$ \\ \hline
3) Phone unlocked, iAuth off & $5.2\%$ \\ \hline
4) Phone unlocked, iAuth on & $7.2\%$ \\ \hline
\end{tabular}
\label{table:power}
\end{table}

\subsubsection{Battery Consumption}
To measure the power consumption, we consider the following four testing scenarios
\begin{enumerate}
\item Phone is locked (i.e., not being used) and iAuth is off
\item Phone is locked and iAuth keeps running
\item Phone is under use and iAuth is off
\item Phone is under use and iAuth is running
\end{enumerate} 
For scenarios 1) and 2), we charge the smartphone battery to $100\%$ and check the battery level after 12 hours. The average difference of the battery charged level from $100\%$ is reported in Table \ref{table:power}. The iAuth-on mode consumes 1.8\% more battery power than the iAuth-off mode each hour. We believe the extra cost in battery consumption caused by iAuth will not affect user experience in daily use.

For scenarios 3) and 4), \emph{the phone under use} means that the user keeps using the phone periodically. During the using time, the user keeps typing notes. The period of using and non-using is five minutes and the test time is 60 minutes. iAuth consumes 2\% more battery power in one hour, which is also an acceptable cost for daily usage.

%% file: conclusion.tex
\section{Conclusions}\label{sec:conclusion}
We have proposed a new re-authentication system, iAuth, to improve the security of a smartphone, and of secret and sensitive data and code in the smartphone or in the cloud accessible through a smartphone. iAuth is an authentication system using multiple sensors built in a user's smartphone, supplemented by auxilliary information from a smartwatch or other wearable device with the same owner as the smartphone. To the best of our knowledge, this is the first work that utilizes both a smartphone and a smartwatch to authenticate the smartphone user. Our system keeps monitoring the users' sensor data and continuously authenticates without any human cooperation. Our system implements fine-grained authentication efficiently and stealthily by using both time and frequency information. 

Experimental results demonstrate the advantage of combining the smartphone and the smartwatch and time-frequency information. iAuth can achieve authentication accuracy up to 92.1\% with negligible system overhead and less than 2\% additional battery consumption. We hope that iAuth can act as a key technique for implicit user authentication in real world scenarios.